\documentclass[conference 10pt]{IEEEtran}
\usepackage[T1]{fontenc}

\ifCLASSINFOpdf
\else
\fi
\usepackage{amssymb,amsbsy,epsfig,float,bm}
\usepackage{graphicx}
\usepackage{dblfloatfix}
\usepackage{subfigure}
\usepackage{balance}
\usepackage{algorithm}
\usepackage[noend]{algpseudocode}
\usepackage{wrapfig,tabularx}
\usepackage[hyphens]{url}
\usepackage[hidelinks]{hyperref}
\hypersetup{breaklinks=true}
\usepackage{algorithm}
\usepackage[noend]{algpseudocode}
\usepackage{subfigure}
\usepackage{float}
\restylefloat{table}
\usepackage{xcolor}
\definecolor{LightCyan}{rgb}{0.88,1,1}
\usepackage{color, colortbl}
\def\BibTeX{{\rm B\kern-.05em{\sc i\kern-.025em b}\kern-.08em
    T\kern-.1667em\lower.7ex\hbox{E}\kern-.125emX}}
\bstctlcite{IEEEexample:BSTcontrol}

\usepackage{amsmath}
\usepackage[cmintegrals]{newtxmath}
\newcommand{\ImgLabel}[1]{\vspace{-3mm}\label{#1}\vspace{-3mm}}
\begin{document}
%
\title{Challenges in Adapting ECH in TLS for Privacy Enhancement over the Internet }
%
%
%




\author{Vinod S. Khandkar, MLiONS Lab, IEOR, IIT Bombay, India\\
Manjesh K. Hanawal, MLiONS Lab, IEOR, IIT Bombay, India \\
Sameer G. Kulkarni, Dept. of CSE, IIT Gandhinagar, India
}

\maketitle

\begin{abstract}
Security and Privacy are crucial in modern Internet services. Transport Layer Security (TLS) has largely addressed the issue of security. However, information about the type of service being accessed goes in plain-text in the initial handshakes of vanilla TLS, thus potentially revealing the activity of users and compromising privacy. The ``Encrypted ClientHello'' or ECH overcomes this issue by extending TLS 1.3 where all of the information that can potentially reveal the service type is masked, thus addressing the privacy issues in TLS 1.3. However, we notice that Internet services tend to use different versions of TLS for application data (primary connection/channel) and supporting data (side channels) such as scheduling information \textit{etc.}. 
Although many internet services have migrated to TLS 1.3, we notice that it is only true for the primary connections which do benefit from TLS 1.3, while the side-channels continue to use lower version of TLS (e.g., 1.2) 
and continue to leak type of service accessed. We demonstrate that privacy information leaked from the side-channels can be used to affect the performance on the primary channels, like blocking or throttling specific service on the internet. Our work demonstrates that adapting ECH on primary channels alone is not sufficient to prevent the privacy leaks and attacks on primary channels. Further, we demonstrate that it is necessary for all of the associated side-channels also to migrate to TLS 1.3 and adapt ECH extension in order to offer complete privacy preservation.
\end{abstract}


%

\section{Introduction}\label{sec:intro}
Security and Privacy are crucial in modern Internet services. 
The widely used Transport Layer Security (TLS)~\cite{rfc2818, the-2021-tls-telemetry-report} 
provides security (data confidentiality and integrity) through the encryption of application data. However, the plain-text fields in TLS handshake messages such as server name indication (SNI)~\cite{rfc3546}, and application layer protocol negotiation (ALPN)~\cite{rfc7301} introduce the concerns over data privacy. Nonetheless, the parameters like SNI and ALPN are critical for proper identification and correct functioning of the network services. Further they also aid the Internet Service Providers (ISPs) to manage the traffic and improve overall network resource utilisation. However, players with malicious intent can use these plain-text information to selectively throttle or block specific services~\cite{wehe_res_2019}  \cite{rjio-site-blocking} resulting in violation of net-neutrality principles~\cite{nn_survey_2020}. This information can also used by malicious actors to profile users' based on their activities. 

The vanilla TLS 1.3 or its lower versions provide no Internet service privacy as they send SNI and ALPN in plain-text. This issue is address by Encrypted-SNI~\cite{rfc8744, cf_esni},  
IPsec \cite{rfc4301} based VPN, Tor \cite{tor}, and  
QUIC protocol~\cite{rfc9000},
by masking the SNI and ALPN fields.  However they have implementations specific issues and
engender several limitations, notably network performance degradation in case of VPN and Tor networks, and plausible security  attacks and lack of forward secrecy with SNI encryption~\cite{rfc8744}. Though  SNI is masked in QUIC, it is retrievable by ISP middle-boxes and thus not providing full privacy. To overcome these issues, an extension is proposed (currently an IETF draft) for the latest version of TLS, i.e., 1.3 to encrypt entire `ClientHello' (ECH) message to mask the server identity in TLS handshake messages~\cite{ietf-tls-esni-14}. 
Fig.~\ref{fig:int_privacy_tax} shows the taxonomy of existing internet protocols and the level of privacy provided by them.
\begin{figure}[!t]
	\centerline{\includegraphics[width=\linewidth]{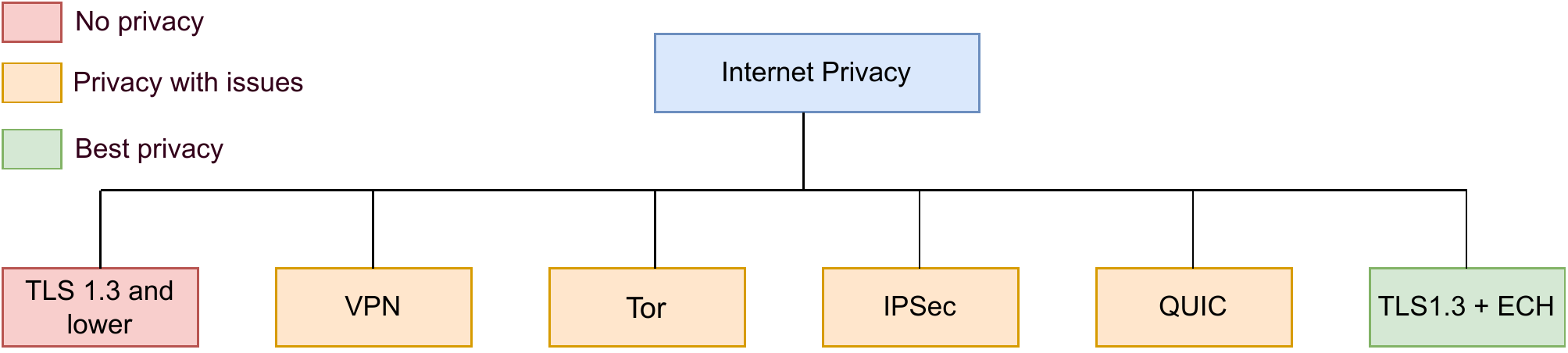}}
	\caption{Existing Internet Security and Privacy taxonomy}
	\ImgLabel{fig:int_privacy_tax}
\end{figure}
Currently, only TLS 1.3 with ECH feature provides the best-in-class privacy to Internet services. However, we argue that the adoption of ECH with TLS 1.3 is still far from truly meeting the privacy challenges. We note that privacy leaks can be exploited by malicious players not only to profile users' activities but also to launch other attacks, \textit{e.g.}, degrading the quality or even blocking access to services.


Modern day internet services dynamically adapt to network conditions to give the best quality and experience to the end users. For example, data rates, frame-sizes, scheduling information are dynamically adjusted based on current congestion levels. For exchange of such information, additional channels are established that are different from the {\em primary channel} used for transfer of service pay-load. The additional channels often carry small amount of intermittent information and we refer to them as {\it side-channels}\footnote{Side-channels are also used to gather user data that could be used for recommendations and commerce.}. By analysing the network logs for given Internet service, we can identify the side-channels for many Internet services (see Sec.~\ref{sec:Side_channels_of_Internet_services}). We list the side-channels for many prominent Internet services such as Primevideo, Hotstar, and Netflix. 
By monitoring traffic of several services, we shown that many services are using lower TLS version for side-channels while they use the latest TLS 1.3 on the primary channels. As ECH is built on TLS 1.3, it can be used to cover privacy issues on the primary channels using TLS 1.3, but not on the side-channels that use lower versions of TLS. Thus, in the current internet, ECH can only partially address the privacy issues as many services still use TLS 1.2 or lower for side-channels.  It is easy to identify a side-channel associated with a service by reading the SNI information that goes in plain-text.  

We demonstrate that partial support of ECH to cover privacy issues in TLS is problematic. As privacy information is still leaked from side-channels due to the lower version of TLS, this information can be use to manipulate the primary channels without identifying them. Specifically, the malicious player can identify the side-channels associated with a service and can block or throttle them which in turn affects the primary-channels as the data transfer on the primary channels rely on the information exchanged on the side-channels. Thus privacy leak in the side-channel can jeopardise the data privacy of Internet services. This loophole allows the Internet Service Provider (ISP) middle-boxes to manipulate the application data delivery without requiring to identifying the primary data channels associated with a given service. The key contributions of our work can be summarized as follows:
\begin{itemize}
    \item We show that most of the Internet services that are currently adopting TLS 1.3 are also using lower version of TLS (TLS 1.2) for side-channels (\textit{e.g.}, scheduling information, ad-insertion, \textit{etc.}) \textit{i.e.}, communication channels other than the primary application data 
    (refer Sec.~\ref{sec:Side_channels_of_Internet_services}).
    \item We experimentally verify that the side channels can reveal the service identity of Internet service using in-house network monitoring tool (refer Sec.~\ref{sec:Side-channels_revealing_service_identity}).
    \item We demonstrate how specific Internet services can be throttled by identifying the side-channels associated with them using commercial traffic-shaper tools (refer Sec.~\ref{sec:tls13_host_identity}).
\end{itemize}

\section{Related work}\label{sec:rel_work}
In a network connection, several fields like SNI and ALPN reveal the identity of service accessed and several solutions exists to mask these fields, namely, the VPN-based solutions providing end-to-end encryption at the network layer (using IPSec \cite{rfc4301}). However, VPN-based solutions are a threat to network performance \cite{vpn_bottleneck}. In Tor \cite{tor}, chained/extended secured connections (series of HTTPS tunnels) are set up through arbitrary intermediate fronting servers until the intended server machine. The key drawback here is the increased workload due to  double encryption and multiple tunneling. 
The SNI obfuscation in QUIC provides the service privacy. However, the ClientHello, that contains SNI, is part of initial packet as a cryptographic handshake message \cite{rfc9001}[Sec.4.3]. Initial packets apply the packet protection process, but use a secret derived from the Destination Connection ID field from the client's first Initial packet and fixed initial\_salt. \cite{rfc9001}[Sec. 5.2] provides the pseudo code to derive the initial secret. As the SNI obfuscation uses the fixed and packet field values, the SNI is completely retrievable as plain-text in QUIC.     

To address privacy of network connections, encryption of the SNI header in TLS was also proposed~\cite{ietf-tls-esni-14,draft-ietf-tls-sni-encryption-09}, and even was adopted by the popular web browser 
Mozilla Firefox \cite{mozilla_esni}.
However, this approach, plagued with several security  vulnerabilities was also discontinued~\cite{esni_issues,cf_esni, firefox,chrome_tls13_downgrade}. The relevant issues are well documented in the internet-draft~\cite{ietf-tls-esni-14,draft-ietf-tls-sni-encryption-09}.
More recently, encryption of the entire ``ClientHello" message in TLS 1.3 has been proposed, named Encrypted ClientHello (ECH).
The details 
are explained in Sec.~\ref{sec:tls13_pri}.

Many machine learning based methods have been developed to identify and associate the Internet services from the active network traffic that operate independent of the protocol parameters~\cite{trcl_survey}. ``AppScanner" \cite{appscanner} is one such tool. 
The tools generate statistical parameters such as (packet size, burst length, source and destination network addresses and port numbers, \textit{etc.}, around a total of 54 different statistical parameters are generated. A simple reinforcement learning based algorithm is then applied/trained for each Internet service using these derived statistical properties or signature.
Even though such techniques achieves the higher accuracy than pure SNI-based techniques, the overhead associated with running these applications is high, primarily due to the use of complex machine algorithms~\cite{trcl_survey}. 
Alternatively, works~\cite{JA3, JARM} have tried to utilize the methods of SSL/TLS fingerprinting using the parameters present in TLS ClientHello and ServerHello exchange messages respectively. However, the reliability of the results depend heavily on the operating system/library versions and custom configurations.  


To the best of our knowledge, our work is the first to study the connection types used by different Internet services and demonstrate the feasibility of privacy attacks on the primary channels (TLS 1.3) even when they could be extended using the privacy preserving ECH protocol. The attack is launched by exploiting privacy information leaked through side channels that do not support ECH. This aspect currently has been left out of scope in the proposed IETF RFC on ECH~\cite{ietf-tls-esni-14}. Further, we also show the vulnerabilities (service throttling/blocking attacks) associated with the partial adoption of ECH.

The paper is organized as follows: In Section \ref{sec:tls13_pri}, we first describe privacy issues in TLS versions 1.2 and 1.3, and how the proposed ECH extension addresses the privacy issue in TLS 1.3. In Section \ref{sec:tls13_pri_leaks}, we present our security and privacy evaluation on variety of the Internet services and their associated primary and side channels with their TLS versions. In Section \ref{sec:tls13_host_identity}, through the commercial grade middlebox (traffic shaper), we demonstrate how the privacy leaked through  side-channels can be used for manipulation and providing differential network services. We give conclusion and future direction in Section \ref{sec:conclusion}.

\section{Privacy in TLS 1.3}\label{sec:tls13_pri}
As of today, TLS (1.2 and 1.3) are the most widely deployed protocol for establishing a secure channel~\cite{rfc8446}.
``HTTP over TLS'' or HTTPS standard adopted the TLS protocol to provide end-to-end secure channel and transmission of encrypted HTTP content. In TLS, a number of privacy-sensitive parameters are negotiated in the clear which leaves a trove of metadata available to network observers, like endpoints' identities and how they use the connections~\cite{coudflare_ech_blog}. We briefly present the background and evolution of TLS 1.2, TLS 1.3 and the associated privacy concerns.

\vspace{-2mm}\subsection{TLS 1.2}\vspace{-2mm}
\begin{figure*}[!ht] 
	\centering
	\subfigure[TLS 1.2 Handshake\label{fig:tls12_hs}]{
		\includegraphics[width=0.3\linewidth]{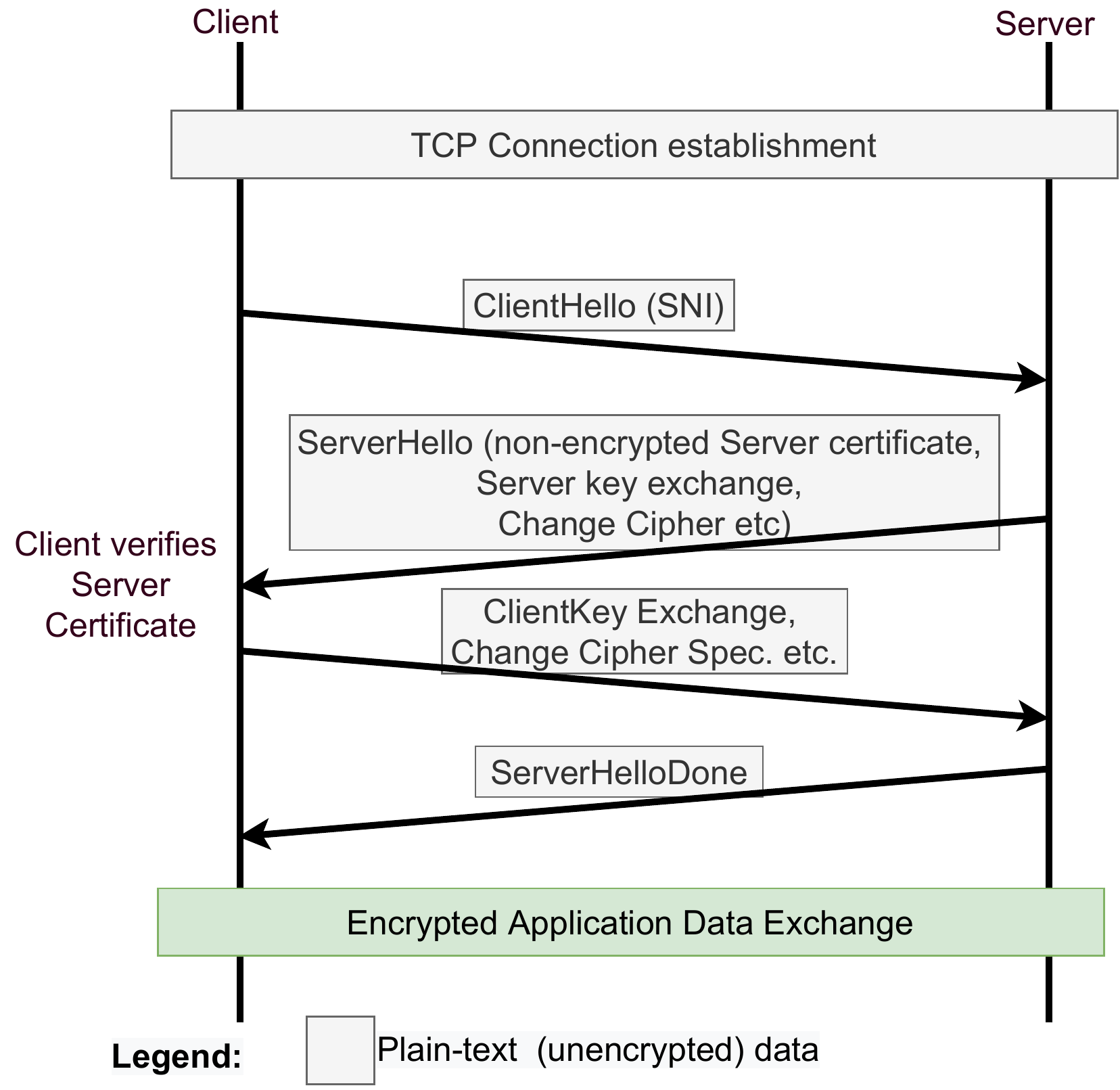}}
	\subfigure[TLS 1.3 Handshake \label{fig:tls13_hs}]{
		\includegraphics[width=0.3\linewidth]{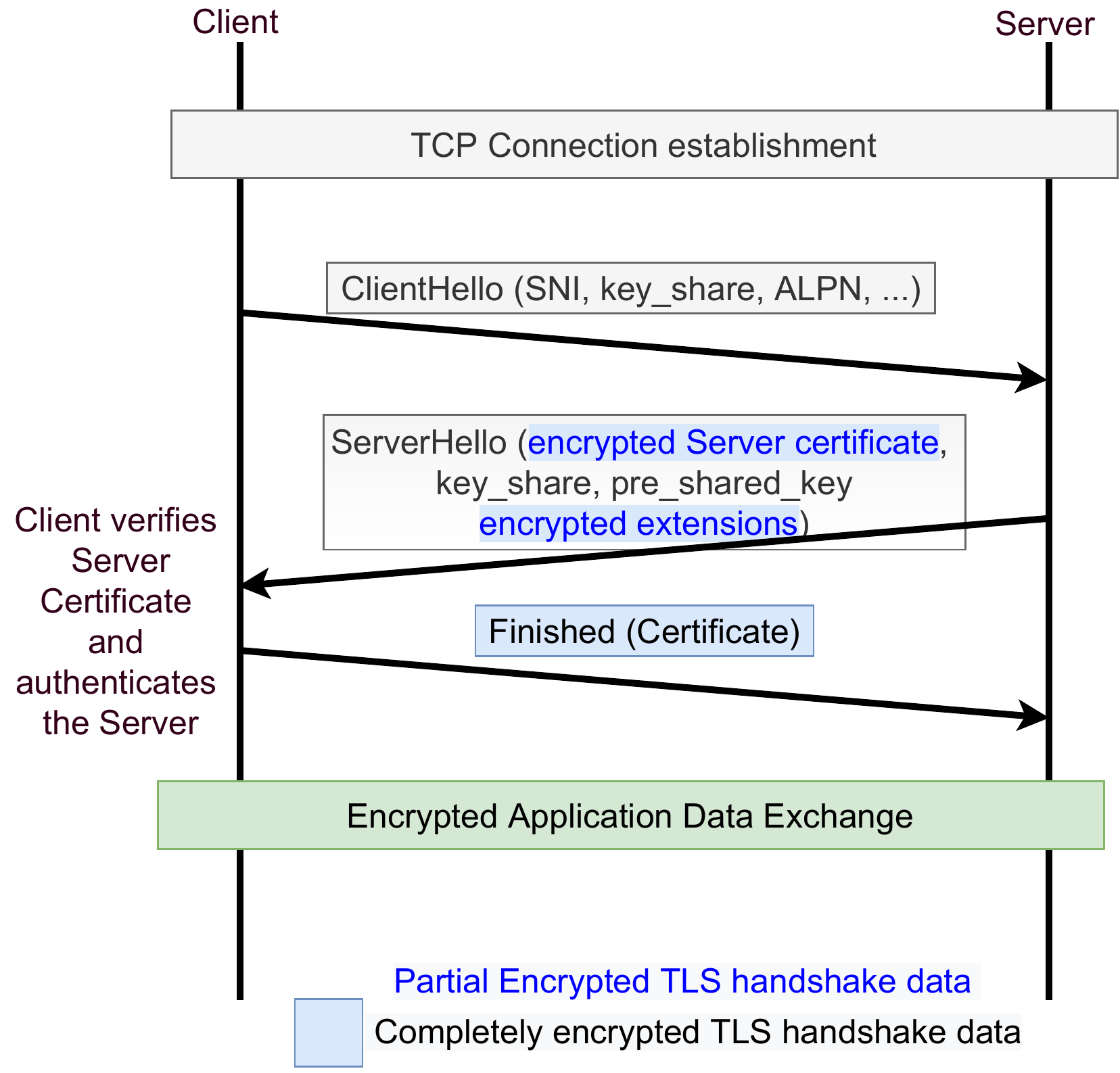}}
	\subfigure[TLS 1.3 + ECH Handshake \label{fig:tls13_ech_hs}]{
		\includegraphics[width=0.3\linewidth]{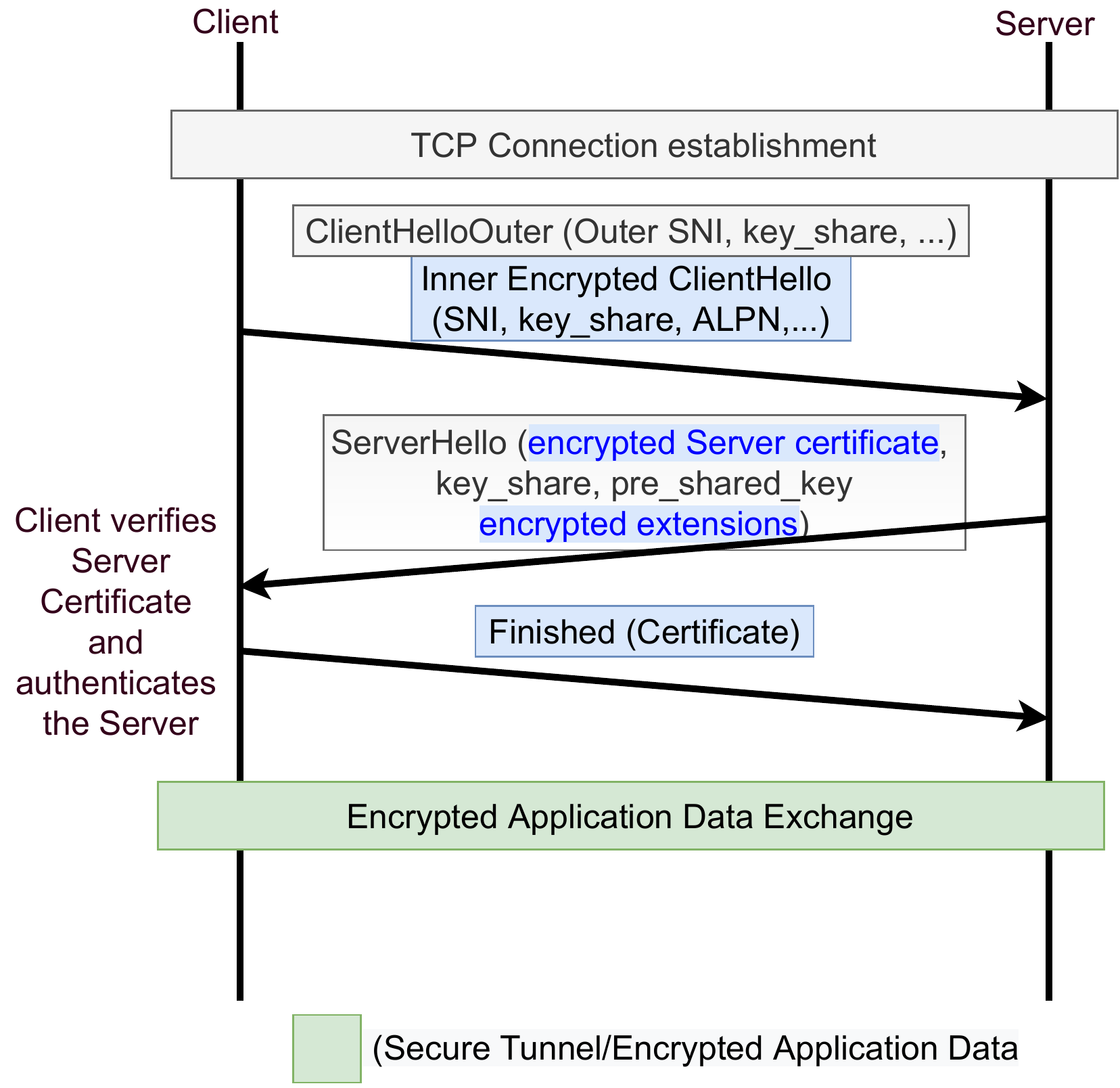}}
	\caption{Message sequence and encryption mode of the messages exchanged during different TLS handshakes (TLS 1.2, TLS 1.3 and TLS 1.3 + ECH).}
	\label{fig:tls12_tls13_hs}
\end{figure*}

Fig.~\ref{fig:tls12_hs} shows the TLS 1.2 handshake that follows the TCP 3-way handshake to negotiate and setup the secure channel. Here, the user-client generates ``ClientHello'' requests using the information from the public URL of the content-server, \textit{e.g., https://www.youtube.com.} This information contains the identity of the server name (SNI) that the client wants to connect with and goes in plain-text ~\cite{rfc3546}. 
Nowadays, it is common for a server (with a single underlying network address) to host multiple `virtual' servers each having their own server certificates. SNI allows the user-client to make distinct secure HTTPS connections towards specific virtual server.
If accepted, the server responds to the user-client request with the ``ServerHello'' message containing the intended server certificate alongside other encryption-related information. 
The client verifies the host-name in the received server certificate to authenticate the server connection, thus helping to thwart any Main-in-the-Middle (MITM) attacks. Moreover, the ``ServerHello'' response from the server to the client carries the Server Certificate and other cipher specifications which are also transferred in the plain-text mode.

Currently, network middle-boxes are well equipped to identify the Internet service using the plain-text SNI information contained in the ``ClientHello'' message.
In-order to avoid this form of privacy leak, encrypted SNI (E-SNI) was proposed where the SNI information in the ``ClientHello'' message is encrypted using the pre-shared  public-key of the server that is made made available in the Domain-Name-Service (DNS) records~\cite{esni}.
However, the E-SNI feature is currently disabled due to various implementation issues~\cite{esni_issues}. 
Further, the middle-boxes can also deduce the service name from the server certificate shared as a part of the TLS ``ServerHello'' message. 
Thus vanilla TLS 1.2 is not able to protect the server host identity over the Internet.

\vspace{-2mm}\subsection{TLS 1.3}\vspace{-2mm}
The latest version of TLS, \textit{i.e.}, TLS 1.3, improves significantly over the TLS 1.2 in terms of performance, security and privacy aspects. 
Figure~\ref{fig:tls13_hs} shows the TLS handshake message exchanges involved in vanilla TLS 1.3. We can observe that in TLS 1.3 the handshake is much more simplified when compared to the TLS 1.2, with reduced number of round-trip communications to finalize the secure communication. Hence, providing better performance. Further, many of the RSA based key exchanges and vulnerable technologies such as the RC4 cipher and CBC-mode ciphers are discontinued. The list of supported cipher specifications have been substantially reduced from 37 in TLS 1.2 to 5 in TLS 1.3, simplifying the usage an enhancing the performance of of TLS 1.3 \cite{rfc8446}.

TLS 1.3 makes use of ``key\_share'' and ``pre\_key\_share'' parameters in the ``ClientHello'' handshake message for encryption purposes. Diffie-Hellman Encryption (DHE) uses the ``key\_share'' parameter to exchange the end-point's public key share required to generate secret key at the remote end-point. The ``pre\_key\_share'' parameter indicates the index of the currently used shared key for encrypting application payload in the list of negotiated shared keys. 
The TLS 1.3 encrypts the server security certificate. 

More specifically, in terms of privacy, it overcomes the plain-text server certificate exchange issue and provides the mechanisms to mask the server's host identity through the the encrypted server certificate. However, the SNI information shared by the client in the ``ClientHello'' message exchange is still shared in plain-text format, engendering the privacy leak.

\vspace{-2mm}\subsection{TLS 1.3 with ECH}\vspace{-2mm}
To remedy this privacy leak through the ClientHello handhshake message, an extension to TLS 1.3, named ``Encrypted Client Hello'' (ECH)~\cite{ech_ietf_draft} has been proposed. 
Figure~\ref{fig:tls13_ech_hs} shows the TLS handshake message exchanges in the proposed ``TLS 1.3 + ECH'' model. 
Here, ECH aims to thwart the privacy leaks and ensure that the client's connection to the servers remain anonymous and indistinguishable over the network. 
In ECH, it is proposed to encrypt and encapsulate the entire legacy ClientHello handshake message (Inner ClientHello), and send it as part of the new ClientHello wrapper (Outer ClientHello) message\footnote{The SNI parameters in the outer clienthello message are set with innocuous values for SNI, ALPN list and other privacy sensitive parameters.}. It uses the public key encryption framework, where the encryption keys (used to encrypt the ClientHello are obtained using the  ``DNS service binding'' \cite{dns_serv_bind} feature. 
There are two modes of exchanging the keys used for the encryption of handshake messages. i) via ''DNS-over-HTTPS'' \cite{rfc8484} information and ii) via a pre-configured key index of the \texttt{ECHConfigId} field within the new ClientHello message.

\begin{table*}[t!]
\caption{Internet service side-channels (SNI and TLS version)}
\label{tab:int_serv_sc}
\begin{tabular}{|c|l|}
\hline
Service& Side channel SNI and TLS version\\
\hline
Hotstar&\begin{tabular}{l}
     hesads.akamaized.net (TLS 1.2),hotstar.com (TLS 1.2), https://img1.hotstarext.com (TLS 1.2),\\
     service.hotstar.com (TLS 1.2),persona.hotstar.com (TLS 1.2),api.hotstar.com (TLS 1.2),\\secure-media.hotstar.com (TLS 1.2),bifrost-api.hotstar.com (TLS 1.2)
\end{tabular} \\ 
\hline
Primevideo &\begin{tabular}{l}
cloudfront.xp-assets.aiv-cdn.net (TLS 1.2),atv-ps-eu.primevideo.com (TLS 1.2),\\m.media-amazon.com (TLS 1.2),fls-eu.amazon.com (TLS 1.2),unagi.amazon.com (TLS 1.2)
\end{tabular}\\ 
\hline
YouTube&\begin{tabular}{l}
fonts.gstatic.com (TLS 1.2),yt3.ggpht.com (TLS 1.2),i.ytimg.com (TLS 1.2),\\pagead2.googlesyndication.com (TLS 1.2)
\end{tabular}\\
\hline
\end{tabular}
\end{table*}

\begin{figure*}[htbp]
	\centerline{\includegraphics[scale=0.35]{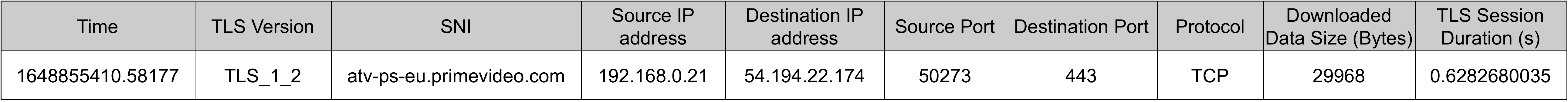}}
	\caption{TLS connection information for Amazon service carrying streaming data over secure channel using TLS 1.3.}
	\label{fig:tls_sc_amazon}
\end{figure*}

\begin{figure}[!ht]
	\centerline{\includegraphics[scale=.4]{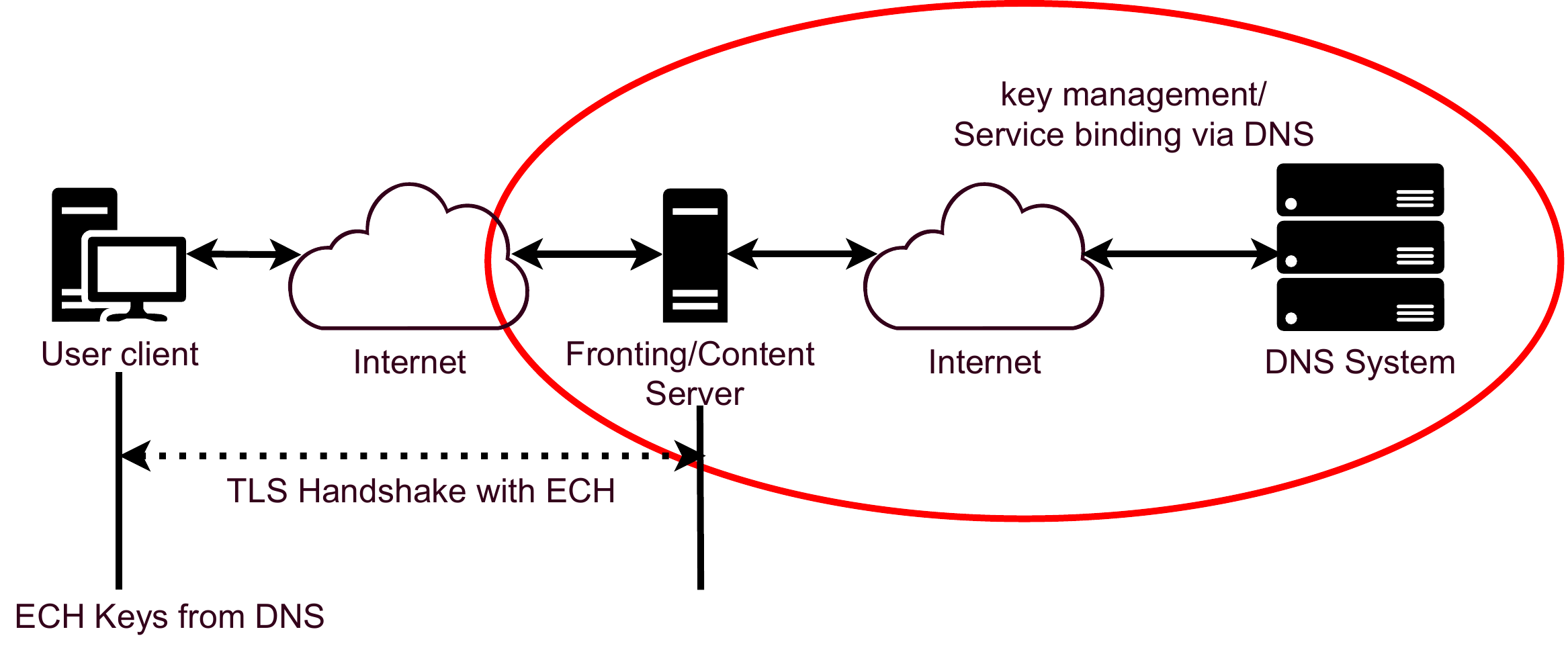}}
	\vspace{-2mm}
	\caption{ECH key management: User client using the encryption key from DNS to encrypt ClientHello message. The DNS and fronting server manages the encryption key for making it correctly available to user-client and avoid any malicious attacks}
	\label{fig:ech_concept}
	\vspace{-2mm}
\end{figure}

\begin{figure*}[!ht] 
	\subfigure[TLS 1.2 \label{fig:sh_tls12}]{
		\includegraphics[scale=0.085]{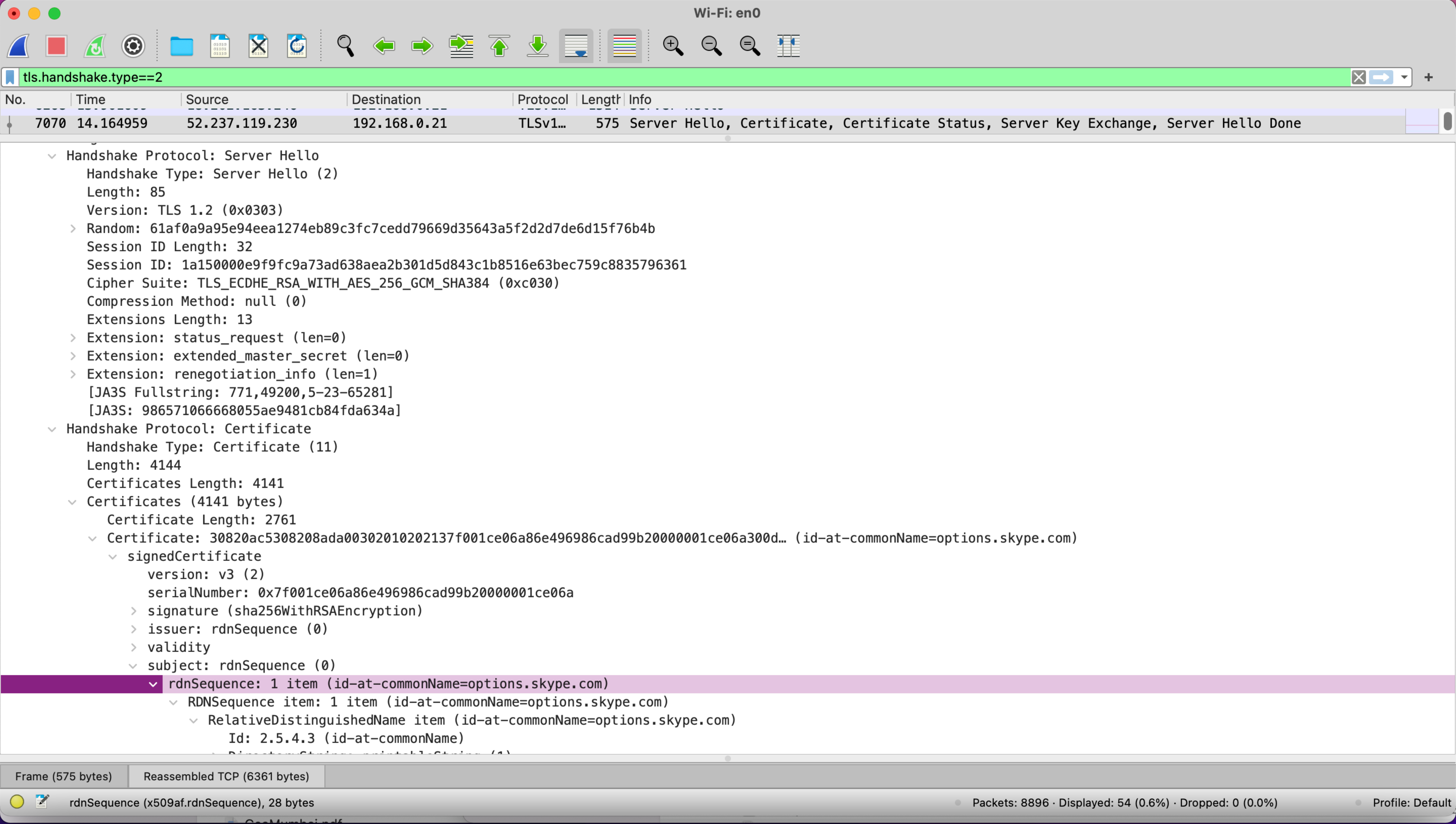}}
	\subfigure[TLS 1.3 \label{fig:sh_tls13}]{
	\includegraphics[scale=0.085]{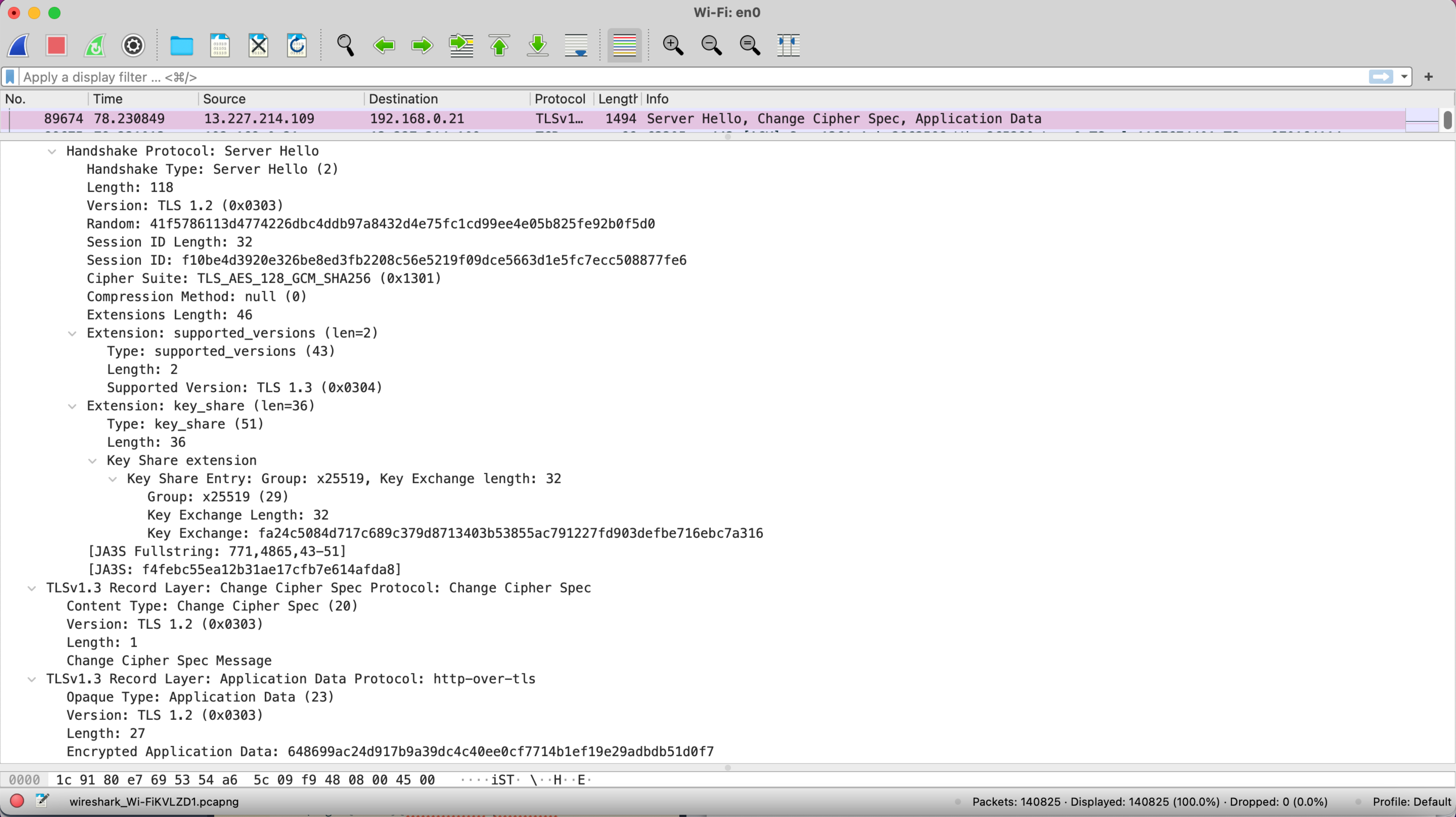}}
	\vspace{-2mm}
	\caption{Server certificate in TLS ServerHello handshake message :Certificate as plain-text in TLS 1.2 while $key\_share$ field exchanged in TLS 1.3 to generate security keys required to encrypt certificate}
	\label{fig:sh_tls12_tls13}
\end{figure*}

Fig.~\ref{fig:ech_concept} shows the key management operation for exchanging ClientHello encryption key using ``DNS-over-HTTPS'' protocol. As shown, the required encryption key is exchanged using SVCB records or HTTPS RR records. The ``Service binding via DNS'' feature adds these records on DNS protocol between user-client and DNS server. This information binds the server's IP address, encrypted ECH public key, and any other connection-specific metadata together and is sent to the user-client whenever DNS information is requested. Note that the fronting server stores the private key part of the ECH encryption key, and it may be exchanged between the fronting server and DNS depending on the agreed terms. The user-client uses the ECH public key to encrypt the ClientHello message. Only the intended fronting server can decrypt this encrypted message with the already stored private part of the ECH key. Moreover, as the fronting server uses this ECH private key part to encrypt the server certificate in the ServerHello handshake message (refer Fig.~\ref{fig:sh_tls12_tls13}), the  confidentiality of the public part of the ECH key is also mandatory. This requirement is fulfilled by using DNS-over-HTTPS feature on user-client to DNS interface, 
Thus TLS 1.3 with ECH provides the data privacy along with confidentiality. 


\section{Privacy leaks in TLS 1.3 with ECH} 
\label{sec:tls13_pri_leaks}
In this section we demonstrate that when an Internet service is accessed, in addition to a primary/main network connection (primary channel) for data transfer, multiple other network connections with different service specific side channels are established between the client and server(s). These additional channels mostly use TLS 1.2 or lower versions that cannot support ECH, and hence can easily leak privacy information. Further, the information leaked by such side-channels can be indirectly used to attack the service offered on the primary-channel even when the primary channels use TLS 1.3 with ECH where no privacy is leaked.

\vspace{-3mm}\subsection{Side-channels of Internet services}
\label{sec:Side_channels_of_Internet_services}\vspace{-2mm}
Many modern services use supporting channels to exchange service specific information. For example, information required to support Dynamic Adaptive Streaming over HTTP (HTTP). These channels facilitate data transfer on primary channel such that the end users experience better quality. We refer to these supporting channels are side-channels. Often a small amount of data is exchanged on the side-channels and hence small overheads are desired on them. Compared to TLS 1.3, TLS 1.2 has low overheads requirements during the initial handshake (no DHE needs to be performed in TLS 1.2), and hence many services continue to use TLS 1.2 on the side-channels. Overheads are not a concern on the primary-channel used for service payload and many services use  TLS 1.3 for the primary-channel. We observed this behaviour on few prominent Internet services such as Amazon prime video,  Hotstar, and YouTube video streaming services. As primary -channels use TLS 1.3, ECH can be used to mask host identify, but on side-channels with TLS 1.2, the host identity remains exposed due to plain-text SNI filed. By analysing the HTTP commands for specific Internet service, we collected service-specific side-channels for popular services as shown in Table~\ref{tab:int_serv_sc}. 

\vspace{-3mm}\subsection{Side-channels revealing service identity}
\label{sec:Side-channels_revealing_service_identity}\vspace{-2mm}
As side-channels use TLS 1.2, the SNI can be read to infer which service they are associated with. We wrote a  script \footnote{https://github.com/MLiONS/ECH-ENSI.git} to extract the source-destination IP address, port numbers, TLS version, SNI and exchanged data size from the network logs recorded for a given Internet service in pcap format. The script compiles the extracted information in CSV format.
Fig.~\ref{fig:tls_sc_amazon} shows the snippet of the extracted CSV format information for the Amazon Primevideo service. As seen from the table, SNI for two channels have domain name primevideo.com indicating that they are the side channels associated with the Primevideo. Further, the script identifies that these side channels use TLS 1.2. 
This experiments confirms that side-channels are still using lower version of TLS and thus their service association can be easily identified.

\section{Traffic manipulation using privacy leaks in TLS 1.3 with ECH}\label{sec:tls13_host_identity}

\begin{figure}[!ht]
	\centerline{\includegraphics[scale=.4]{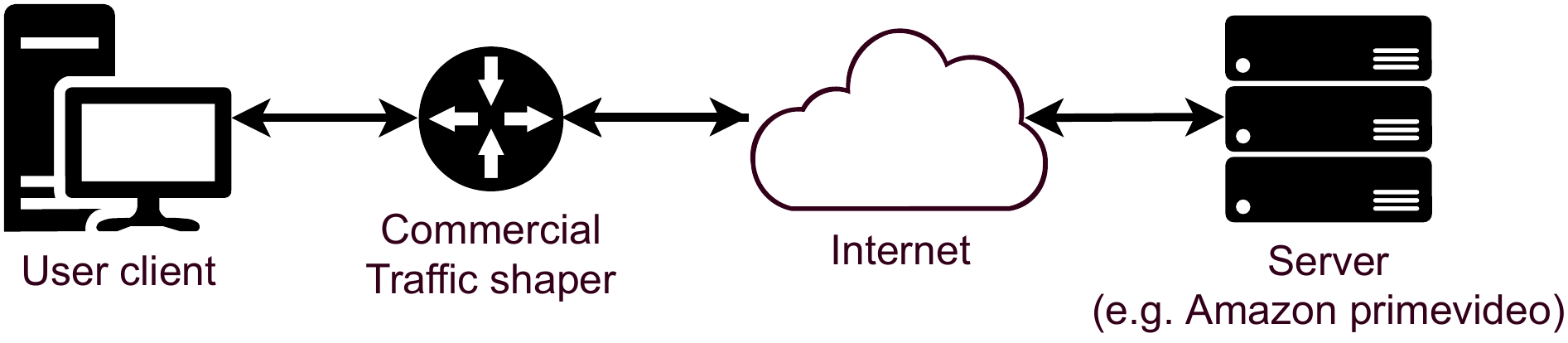}}
	\caption{TLS 1.3 side-channel experimental setup}
	\label{fig:tls13_exp_setup}
\end{figure}

\begin{figure*}[!ht] 
	\centering
	\subfigure[Hotstar\label{fig:hs_normal}]{
		\includegraphics[width=0.4\linewidth]{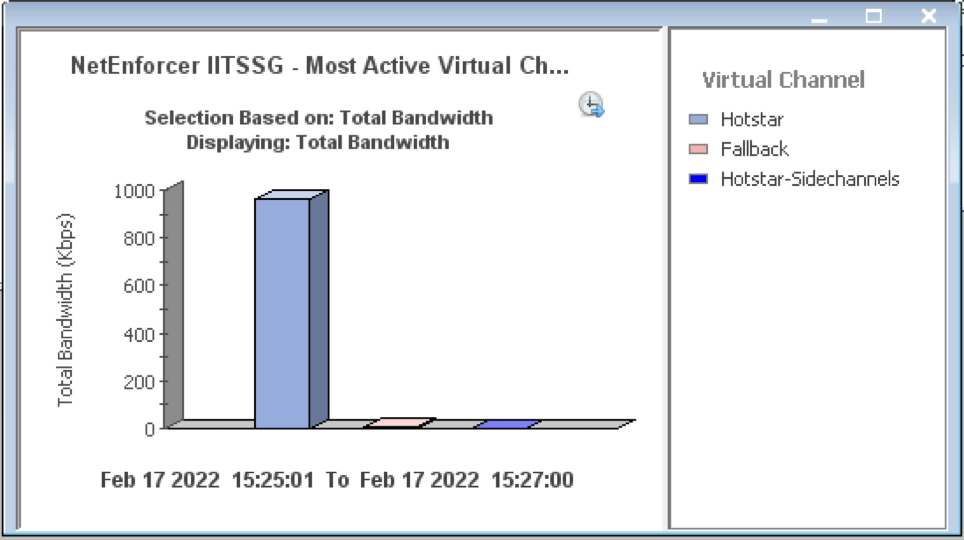}}
	\subfigure[Amazon Primevideo\label{fig:pv_normal}]{
		\includegraphics[width=0.4\linewidth]{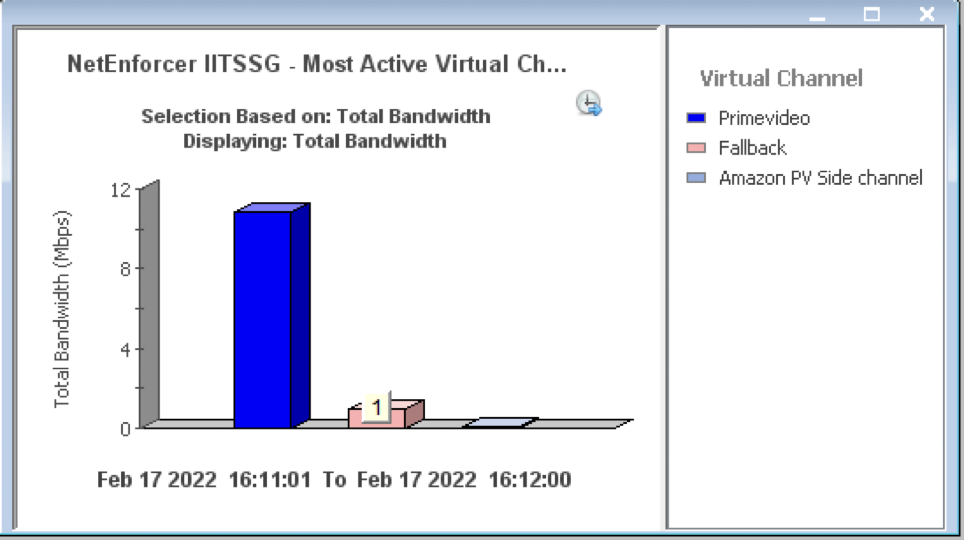}}
	\caption{Normal operation of Hotstar and Primevideo when their side channels are not affected.}
	\label{fig:hs_pv_normal}
	\vspace{-2mm}
\end{figure*}

\begin{figure*}[!ht] 
	\centering
	\subfigure[Browser window (No video)\label{fig:hs_ts}]{
		\includegraphics[width=0.4\linewidth]{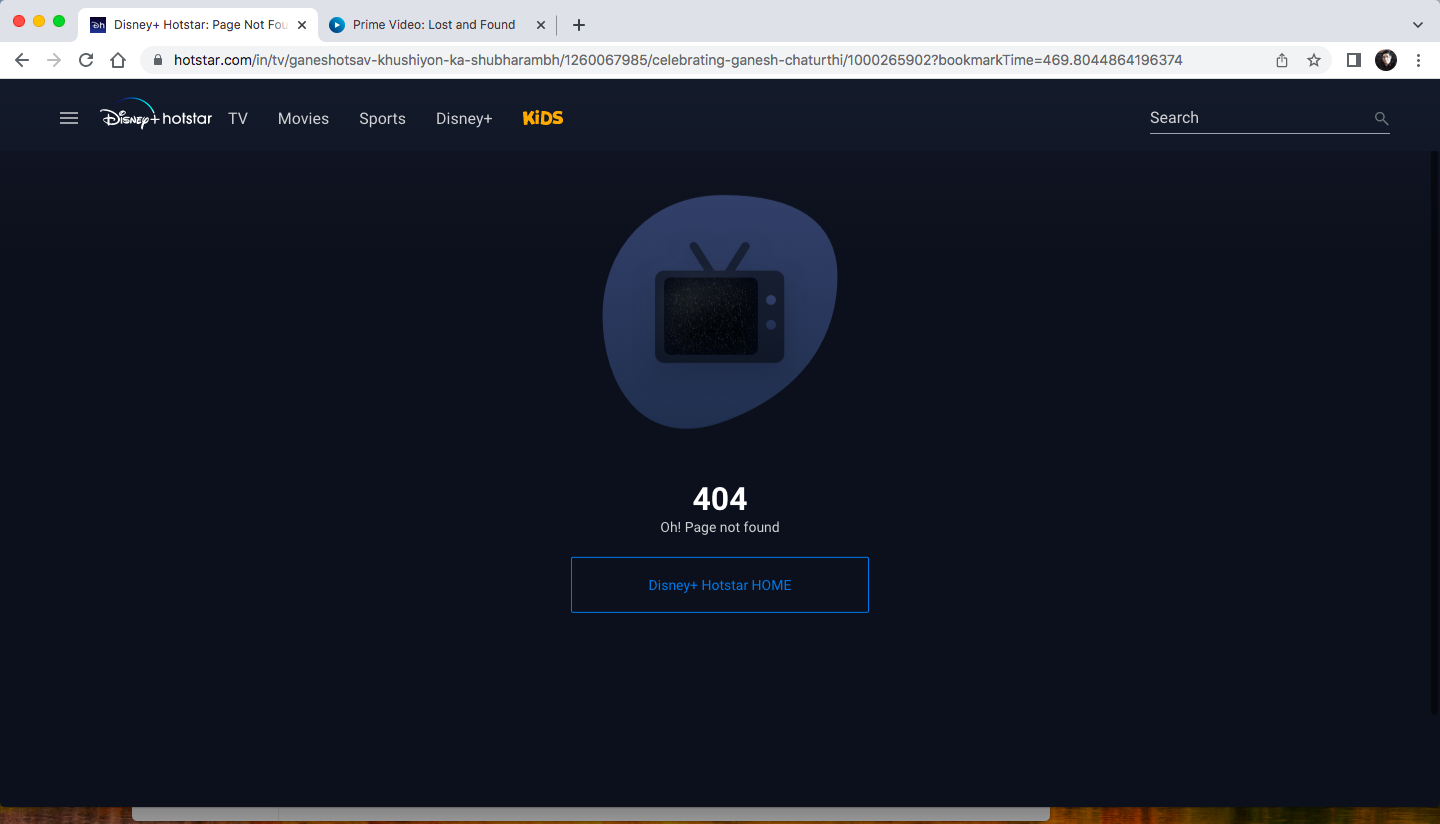}}
	\subfigure[Traffic shaper outcome (low data rate Hotstar download)\label{fig:hs_bw}]{
		\includegraphics[width=0.4\linewidth]{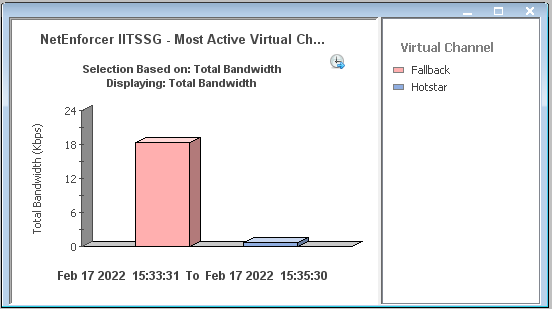}}
	\caption{Hotstar: Effect of blocking the side-channels in Hotstar either before the start of video or during the ongoing video playback session.}
	\label{fig:hs_ts_bw}
\end{figure*}

\begin{figure*}[!ht] 
	\centering
	\subfigure[Browser window (Video error)\label{fig:pv_ts}]{
		\includegraphics[width=0.4\linewidth]{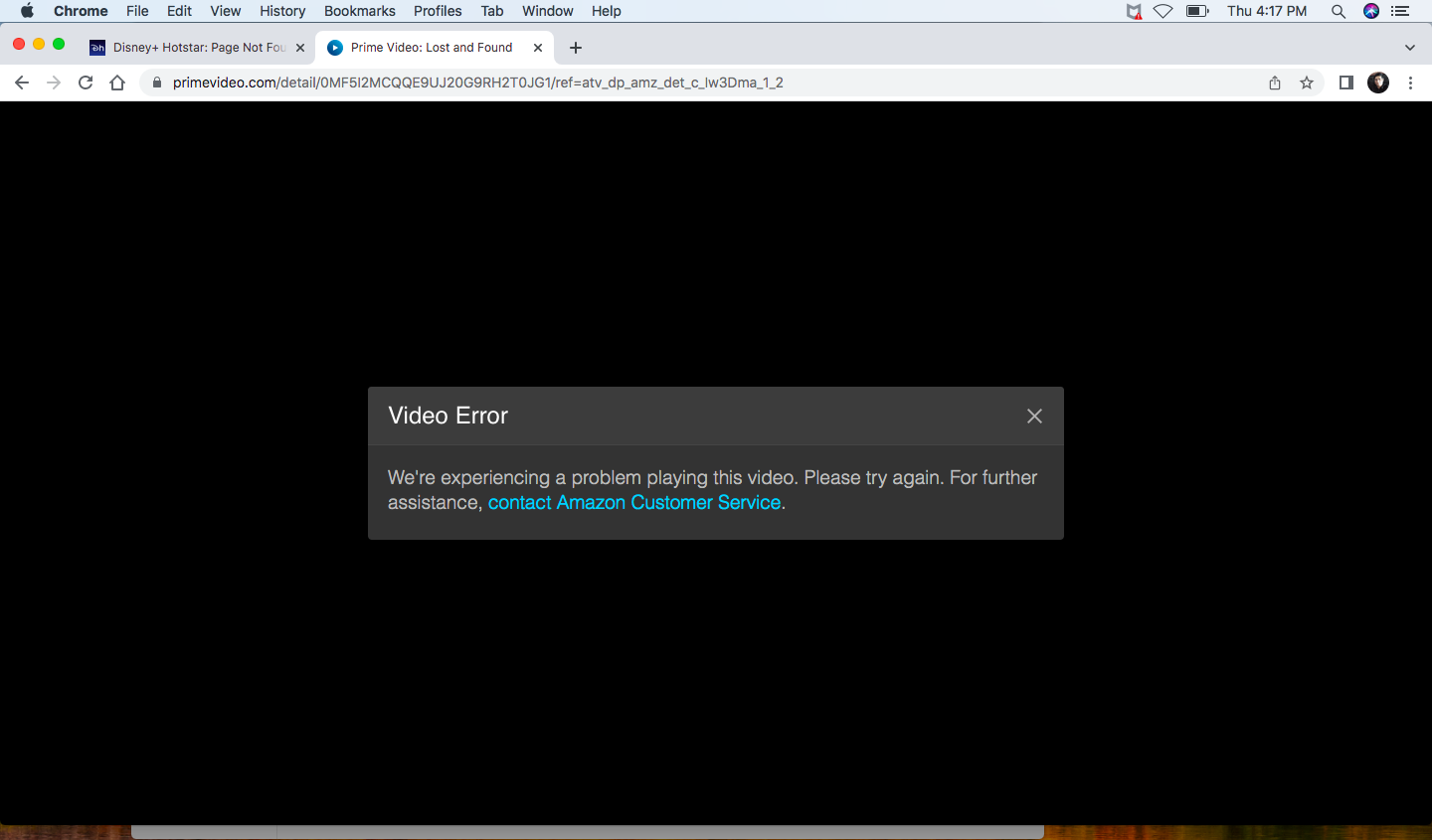}}
	\subfigure[Traffic shaper outcome\label{fig:pv_bw}]{
		\includegraphics[width=0.4\linewidth]{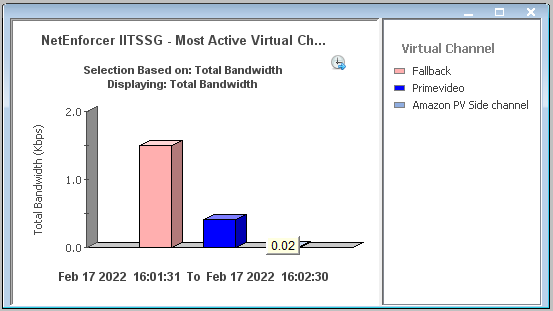}}
	\vspace{-2mm}
	\caption{PrimeVideo: Side-channels throttled before starting video}
	\label{fig:pv_ts_bw}
\end{figure*}

\begin{figure*}[!ht] 
	\centering
	\subfigure[Browser window (Lower quality video playback)\label{fig:pv_ts_d}]{
		\includegraphics[width=0.4\linewidth]{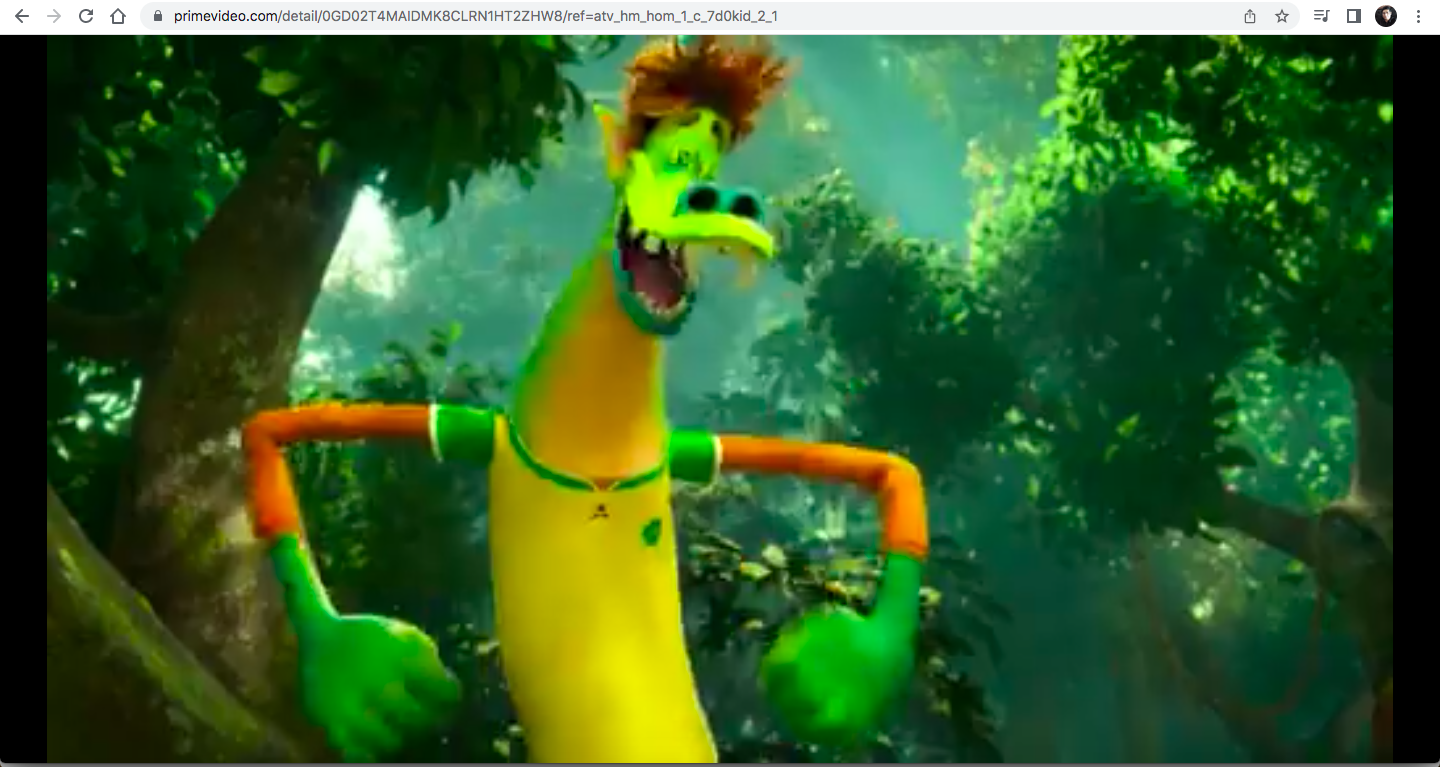}}
	\subfigure[Traffic shaper outcome\label{fig:pv_bw_d}]{
		\includegraphics[width=0.4\linewidth]{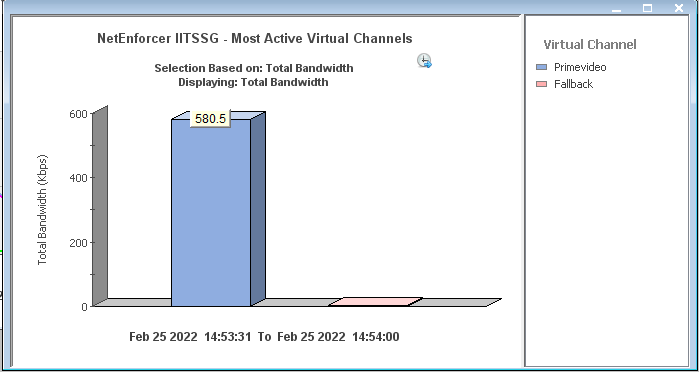}}
	\vspace{-2mm}
	\caption{PrimeVideo : Side-channels blocked during video}
	\label{fig:pv_ts_bw_d}
\end{figure*}
\begin{figure*}[!ht] 
	\centering
	\subfigure[Missing recommended video thumbnails/advertisements in the browser window \label{fig:yt_ts}]{
		\includegraphics[width=0.4\linewidth]{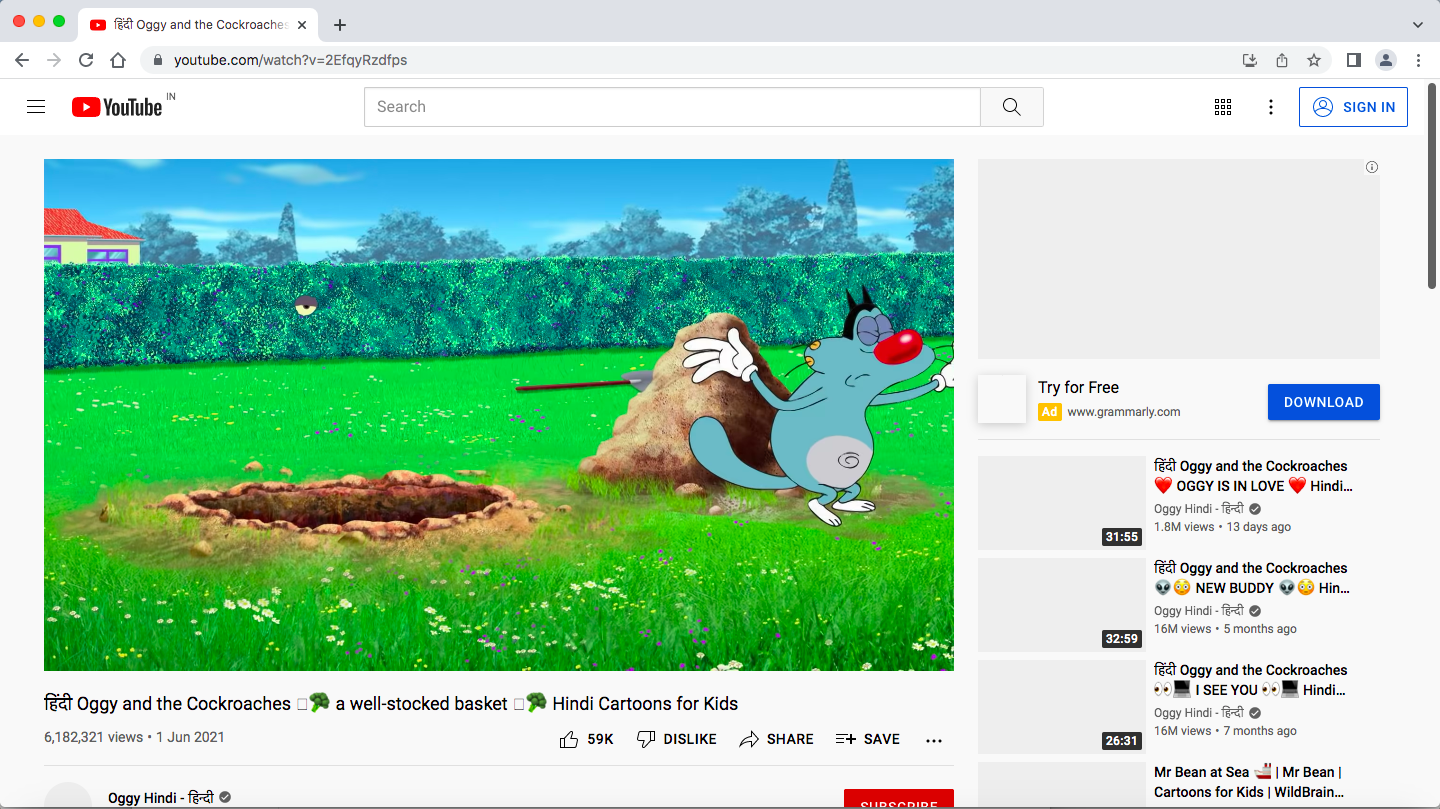}}
	\subfigure[Traffic shaper outcome\label{fig:yt_bw}]{
		\includegraphics[width=0.4\linewidth]{Figs/PVSDSidechannelBlockedDuringVideo.png}}
	\caption{YouTube : Side-channels blocked during video}
	\label{fig:yt_ts_bw}

\end{figure*}
\begin{table*}[!t]
    \renewcommand{\arraystretch}{1.4}
    \vspace{-2mm}
    \caption{Impact of side channel blocking on streaming service video playout}
    \label{tab:tls13_sc}
    \centering
    \small
	\begin{tabular}{|c|c|c|}
		\hline
		Service & Before Video playout & During video playout \\
		\hline
		Hotstar & No video & No Video  \\ 	\hline
		Primevideo & No Video & Video Playout : Reduced rate and quality downgrade \\ \hline
		YouTube & Video playout, no thumbnails on webpage & Video playout, no thumbnails on webpage\\
		\hline
	\end{tabular}
\end{table*}

As described in the previous section, if SNI field can be read for a channel, it can be associated with a service type. Hence, it is possible for all the side-channels using TLS 1.2 to be associated with the corresponding primary channel (service) by reading their plain-text SNI field. 
However, when the primary-channel of a service uses TLS 1.3, and is integrated with ECH\footnote{Currently the SNI information in TLS 1.3 is in plain-text format. However, when ECH gets integrated with TLS 1.3, we would not able to decode the SNI information.}, the SNI parameter remains encrypted and it is not possible to associate it to a specific service. Note that, on the contrary, when a series of connections (Primary channel using TLS 1.3 + ECH and others using TLS 1.2) are established, it becomes easier to associate the service type of a primary channel (TLS 1.3 + ECH) even when it is encrypted due to the unique presence of secondary channels. We exploit this feature to successfully target and identify the type of service associated with primary channel even without looking at the SNI headers for TLS 1.3 (\textit{i.e.} effectively treating TLS 1.3 channel as ECH enabled TLS 1.3). 

In this section, we experiment and analyze the dependence and impact on the performance of a service (primary channel) when we throttle/block the identified side-channels.
For our study and analysis, we use three different streaming services namely - Amazon primevideo, Hotstar and YouTube. 
We used a commercial traffic shaper to block/throttle the traffic on the secondary channels. Our experimental set up is as shown in Fig.~\ref{fig:tls13_exp_setup}. 
We obtained side-channel associated with the services by looking for connections using TLS 1.2 and reading their SNIs from the traffic log for given streaming service. We used traffic shaper from Allot system (Allot SSG200 for Visibility \& Control)\footnote{Allot Traffic Shaper: \url{ https://www.allotworks.com/SSG-200.asp}}, to record the Internet data activity on various user-client connections for a given streaming service e.g. streaming data channel or side-channel. The traffic shaper has the provision to block/throttle a given list of SNIs. We analyzed the effect of blocking side-channels for the following scenarios:
\begin{itemize}
    \item side-channels blocking/throttled before starting of web service e.g. before starting video
    \item side-channels blocking/throttled during web service is ON e.g. during the ongoing video playback
\end{itemize}

\subsection{Hotstar video streaming service}
In this section, we demonstrate the method to manipulate and affect the traffic performance for Hotstar steaming service.
\subsubsection{Normal scenario}
The traffic shaper user-interface outcome for normal operation of Hotstar service is shown in Fig.~\ref{fig:hs_normal}. It shows Hotstar service's streaming data, side channel data activity, and activities from all other background applications (named Fallback). We used free subscription version of Hotstar where only SD quality is allowed with peak data rate of around 1 Mbps. The side-channel activity is prominent at the beginning of the video play when video download occurs at relatively lower speed of 500-600 Kbps.
\subsubsection{Side-channels blocked before starting/during video play}
As shown in Fig.~\ref{fig:hs_ts_bw}, when side channels are blocked for Hotstar service at the beginning of the video, the video playout stops (refer Fig.~\ref{fig:hs_ts}) by displaying error message to user. Note that no data activity is visible in traffic shaper record (refer Fig.~\ref{fig:hs_bw}) for side- channels confirming that they are successfully blocked and the data activity on the primary-channel is almost nil. When the side-channels are blocked while the video play is in progress, the video playback stops after sometime but not immediately. We suspect this behavior could be for the reason that the Hotstar might not be downloading the entire schedule for video in the beginning and continue to access the payload on the primary channel. The video stops when player needs the new schedule but cannot obtain it due to the blocked side-channels.

\subsection{Amazon Primevideo}
In this section we demonstrate the method to manipulate and affect the traffic performance for Amazon Primevideo service.
\subsubsection{Normal operations}

Fig.~\ref{fig:pv_normal} shows the traffic shaper outcome for normal operation of Primevideo service. We used paid subscription of Primevideo which gave data rates of around 11 Mbps. The side-channel activities are prominent even at the peak download rates. It can be attributed to periodic downloading of supporting data e.g. scheduling information on side-channels. 

\subsubsection{Side-channels blocked before staring/during video play}

When side channels are blocked for Primevideo service before the video-play begins, the Primevideo player cannot download the schedule for the video play. Hence the video player throws error as shown in Fig.~\ref{fig:pv_ts} while traffic shaper records low data activity for Primevideo service as shown in Fig.~\ref{fig:pv_bw}. When side-channels are blocked while the video play is in progress, the video playout does not stop as shown in Fig.~\ref{fig:pv_ts_d} even though no data activity for the side-channel is observed as seen from Fig.~\ref{fig:pv_bw_d}. Moreover when monitoring HTTP activity through Chrome’s network analyser reveals that the Amazon primevideo changes its downloads to a new server with reduced rate and quality as shown in Fig.~\ref{fig:pv_bw_d}. Note that video plays with its highest quality when side-channels are not blocked, but switches to low quality when the side-channels are blocked\footnote{Amazon Primevideo has three distinct qualities/resolutions of video playback namely Good (480p SD quality, ~0.38GB per hour), Better (720p HD quality, 1.4GB per hour) and Best (1080p Full HD quality, 6.84GB per hour)}. 
If side-channels associated with the new data server are further blocked, no video is played, i.e., complete blocking of Primevideo is accomplished without blocking primary channel to the data server.

\subsection{YouTube video streaming}
When side channels are blocked during or before starting YouTube video, the video continues to play with 4—5 Mbps data download speed as shown in Fig.~\ref{fig:yt_ts_bw}. However, the thumbnails for recommended videos or advertises are not visible in the browser window (refer Fig.~\ref{fig:yt_bw}). This behaviour implies that YouTube is fetching only thumbnail/advertisement material through side-channels and information related to play the main content is fetched from the primary-channel itself.

Table~\ref{tab:tls13_sc} summarises the effect of blocking side-channels on performance of streaming services.

\subsection{Other Internet services}

We conducted experiments on other type of services like audio streaming (e.g. Amazon prime music) and news content (e.g. BBC), where by blocking the side-channels some information displayed on the webpage are lost., for example advertisements and thumbnails. We skip these details due to space limitations.

\section{Discussion and future work}\label{sec:discussion}
As our experiments demonstrated, many popular services use TLS 1.2 for side-channels which makes them easy to identify and associate with a service. Though it is not possible identify primary-channel associated with a service when its SNI is masked, one can identify a channels as primary-channel of a streaming service
as they carry large amount of data (in the order of MBs) compared to the relatively smaller data (in the order of KBs) carried by the side-channels. As long as a malicious attacker can identify a side-channel associated with a service, it can attack its primary-channel without requiring to identify it. However, the distinction between primary-channel and side-channel may not be clear for other Internet services such as online newspaper or blog sites as data carried on the primary-channel for these services could be of similar amount as carried on the other side-channels. However, the primary-channels in such services can still be distinguished from side-channels
by monitoring more metrics' such as TLS session length. As session lengths are expected to be larger on the primary-channels compared to the secondary-channels, they can be a reliably used as a metric to differentiate the  side-channels for the primary channels. We note that it is not required that the primary-channels needs to be identified to effect a services. It is enough to identify the side-channels associated with a service and that is easy to do as most of the side-channels use vanilla TLS which exposes SNI and reveals service/host identity.   

Thus we believe that data privacy issues is not solved by addressing it on only TLS 1.3. Still a significant portions of Internet traffic is using TLS 1.2 and the privacy in them also needs to be addressed. As demonstrated by our experiments, when a service is accessed multiple connections are established all of which are not using TLS 1.3. Thus any solution that addresses the privacy issues in only TLS 1.3 is only providing partial privacy which could be easily exploited to attack the channels which have full privacy protection (with ECH).
In E-TLS in \cite{etls}, in order to remedy the privacy leaks with SNI, the authors proposed a novel SNI masking method.
E-TLS uses 2-phase TLS handshake to encrypt entire TLS handshake and can work for both TLS 1.2 and 1.3. However, E-TLS needs the fronting servers to respond in a specific way which can potentially cause deployment challenges. Hence, we argue that providing complete privacy in the current Internet is still an open problem and requires further attention.

\section{Conclusion}\label{sec:conclusion}
This paper demonstrated that the new extension to TLS 1.3, named Encrypted Client Hellos (ECH) does not provide the intended privacy. This is primarily because many service still uses  TLS 1.2 while ECH only works with TLS 1.3. Many services on Internet use side-channels with low overheads to provide better experience for users on the primary-channel used for actual service pay-load. The side-channel mostly use TLS 1.2 which reveal identity of the service which can be used to attack primary-channel even though they use ECH and their identify is not known. 
We demonstrated this type of attacks on several services, by first identifying side-channels associated with the services through their SNIs and then using this information to attack the primary-channels. 
Thus, our work demonstrates the core privacy challenges in the current deployment of TLS 1.3 and ECH. We alarm that a limited transition to TLS 1.3 + ECH may not achieve the desired goals of privacy and anonymity, and fail to protect from malicious attacks that throttle/block specific internet services, until and unless sufficient care is taken to migrate all the associated connections (secondary channels using TLS 1.2) to TLS 1.3 and adopt ECH universally for all the connections.



%

\ifCLASSOPTIONcaptionsoff
  \newpage
\fi



\bibliographystyle{IEEEtran}
\bibliography{IEEEabrv, vinod_bibfile}

\begin{thebibliography}{10}
\bibitem{firefox}
Cloudflare reports esni is no longer working.
\end{thebibliography}
\end{document}